\documentclass[aps,prl,twocolumn,nofootinbib,longbibliography,superscriptaddress,aip]{revtex4-1}
\usepackage{hyperref}
\usepackage{amsmath}
\usepackage{amssymb}
\usepackage{amsthm}
\usepackage{graphicx}
\usepackage{physics}
\usepackage{color}
\usepackage{mathrsfs}
\usepackage{dsfont}
\definecolor{coloredit}{rgb}{0.0, 0.53, 0.74}

\begin{document}

\title{Microscopic origin of the entropy of astrophysical black holes}

\author{Vijay Balasubramanian}
\affiliation{David Rittenhouse Laboratory, University of Pennsylvania, 209 S.33rd Street, Philadelphia, PA 19104, USA}
\author{Albion Lawrence}
\affiliation{Martin Fisher School of Physics, Brandeis University, Waltham, Massachusetts 02453, USA}
\author{Javier M.~Mag\'{a}n}
\affiliation{David Rittenhouse Laboratory, University of Pennsylvania, 209 S.33rd Street, Philadelphia, PA 19104, USA}
\author{Martin~Sasieta}
\affiliation{Martin Fisher School of Physics, Brandeis University, Waltham, Massachusetts 02453, USA}

\begin{abstract}
We construct an infinite family of microstates for black holes in Minkowski spacetime which have effective semiclassical descriptions in terms of collapsing dust shells in the black hole interior.  Quantum mechanical wormholes cause these states to have exponentially small, but universal, overlaps. We show that these overlaps imply that the microstates span a Hilbert space of log dimension equal to the event horizon area divided by four times the Newton constant, explaining the statistical origin of the Bekenstein-Hawking entropy.
\end{abstract}

\maketitle

\section{Introduction} 

Bekenstein and Hawking \cite{Bekenstein:1973ur,Hawking:1975vcx}  proposed, on the basis of general relativity and quantum mechanics in curved spacetimes, that black holes behave as thermodynamic objects, and carry an entropy  $S = {A / 4 G}$, where $A$ is the area of the event horizon, $G$ is Newton's constant, and we are working in units where Planck's constant and the speed of light are $1$. This remarkable formula is universal. It applies to any black hole regardless of its mass, charge, or angular momentum, and in any spacetime dimension.

What is the origin of this entropy? Statistical mechanics asserts that the thermodynamic entropy of a classical system equals the logarithm of the number of microstates consistent with the macroscopic parameters.  Quantum mechanics complicates matters. Quantum states form a Hilbert space; so any suitably normalized linear combination of microstates is also a microstate.  Thus, in quantum systems we instead identify entropy as the logarithm of the {\it dimension} of the Hilbert space.  To give a statistical mechanical interpretation of black hole entropy, we have to determine the dimension of the underlying quantum gravity Hilbert space describing a black hole.

This fundamental problem was solved in a special case by Strominger and Vafa \cite{Strominger:1996sh} who explained the entropy of certain supersymmetric black holes in terms of the Hilbert space of underlying string theoretic microstates.  These  calculations were possible because the black holes in question: (1) have multiple types of electric and magnetic charges (unlike our world where there is one electromagnetic field, and no magnetic charge); (2) are extremal (unlike most astrophysical black holes) so that the mass achieves a certain lower bound in terms of the charges required for avoiding naked spacetime singularities; and (3) are supersymmetric, in that they retain a fraction of the supersymmetry of the theories in which they are defined (unlike real black holes which have no supersymmetry to break), which is central to the computability of the entropy.  Furthermore, the analysis relied on technical details of the ultraviolet completion of gravity in string theory, which include many extra dimensions and exotic extended solitonic objects of cosmic scale, thereby obscuring the nature of these microstates in the semiclassical description of the black hole. The fundamental question has thus remained: can we give a universal microscopic explanation for the entropy of astrophysical black holes?  Here, we propose an answer to this question.

Briefly, we will exploit the fact that in quantum statistical mechanics, any superposition of microstates is also a microstate, where a microstate is a normalizable vector in the Hilbert space with fixed expectation values for macroscopic observables. Thus, rather than a specific basis of typical black hole microstates, we simply seek any set of states that is large enough to span the entire Hilbert space.  We also require this set to be under sufficient control for us to compute the Gram matrix of state overlaps.   The rank of the Gram matrix determines the maximum number of linearly independent microstates, giving the dimension of the Hilbert space. Equivalently, the logarithm of this rank quantifies the statistical entropy.

To this end, we construct an infinite family of atypical, but well-controlled, microstates for black holes in Minkowski space with effective semiclassical descriptions, which include dust shells in the black hole interior.  Our construction follows from general relativity, and does not require any exotic ingredients.  Astrophysical black holes generally have some angular momentum, but we will focus on non-rotating black holes for analytical simplicity. Extending methods developed in \cite{Balasubramanian:2022gmo} for universes with a negative cosmological constant, we compute the  overlaps of our microstates in quantum gravity.  We find that they span a Hilbert space of dimension precisely equal to the exponential of the Bekenstein-Hawking entropy.  This finding explains the microscopic origin of black hole thermodynamics.

\section{Black Hole Microstates}

\begin{figure}[h]
    \centering
    \includegraphics[width = .4\textwidth]{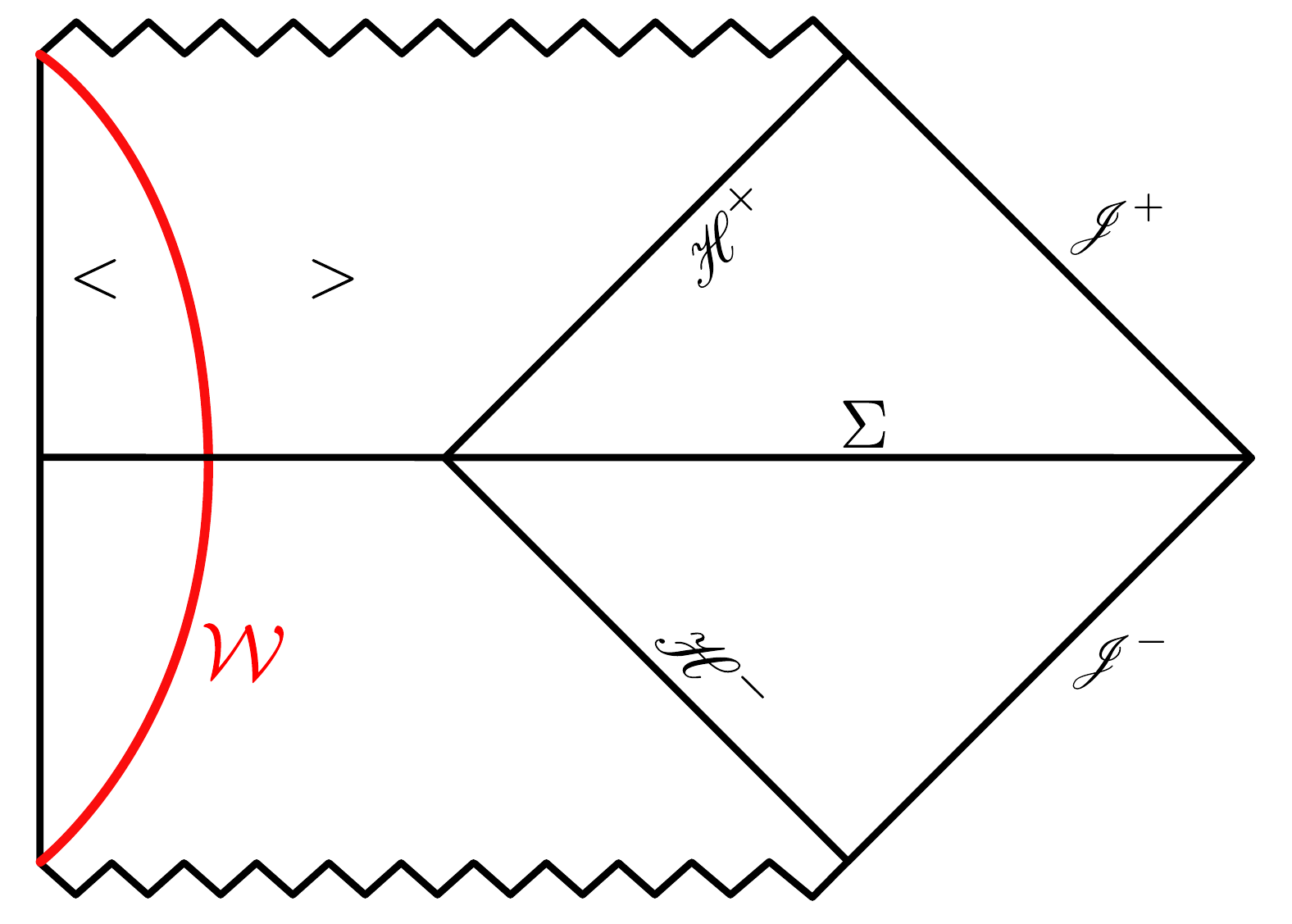}
    \caption{Penrose diagram of the time evolution of a microstate of an eternal one-sided black hole. The semiclassical state is defined at the time reflection-symmetric Cauchy slice $\Sigma$. The exterior geometry extends between the future/past horizons $\mathscr{H}^\pm$ and the conformal null boundaries $\mathscr{J}^\pm$. The interior contains a thin shell $\mathcal{W}$, which divides the geometry between a region of flat space $<$ inside the shell, and a region of black hole geometry $>$ otuside the shell.  The zigazag lines at the bottom and top are the the white hole and black hole singularities where time starts and ends. The semiclassical state on $\Sigma$ is non-singular and perfectly well defined.}
    \label{fig:shell}
\end{figure}

\begin{figure}[h]
    \centering
    \includegraphics[width = .45\textwidth]{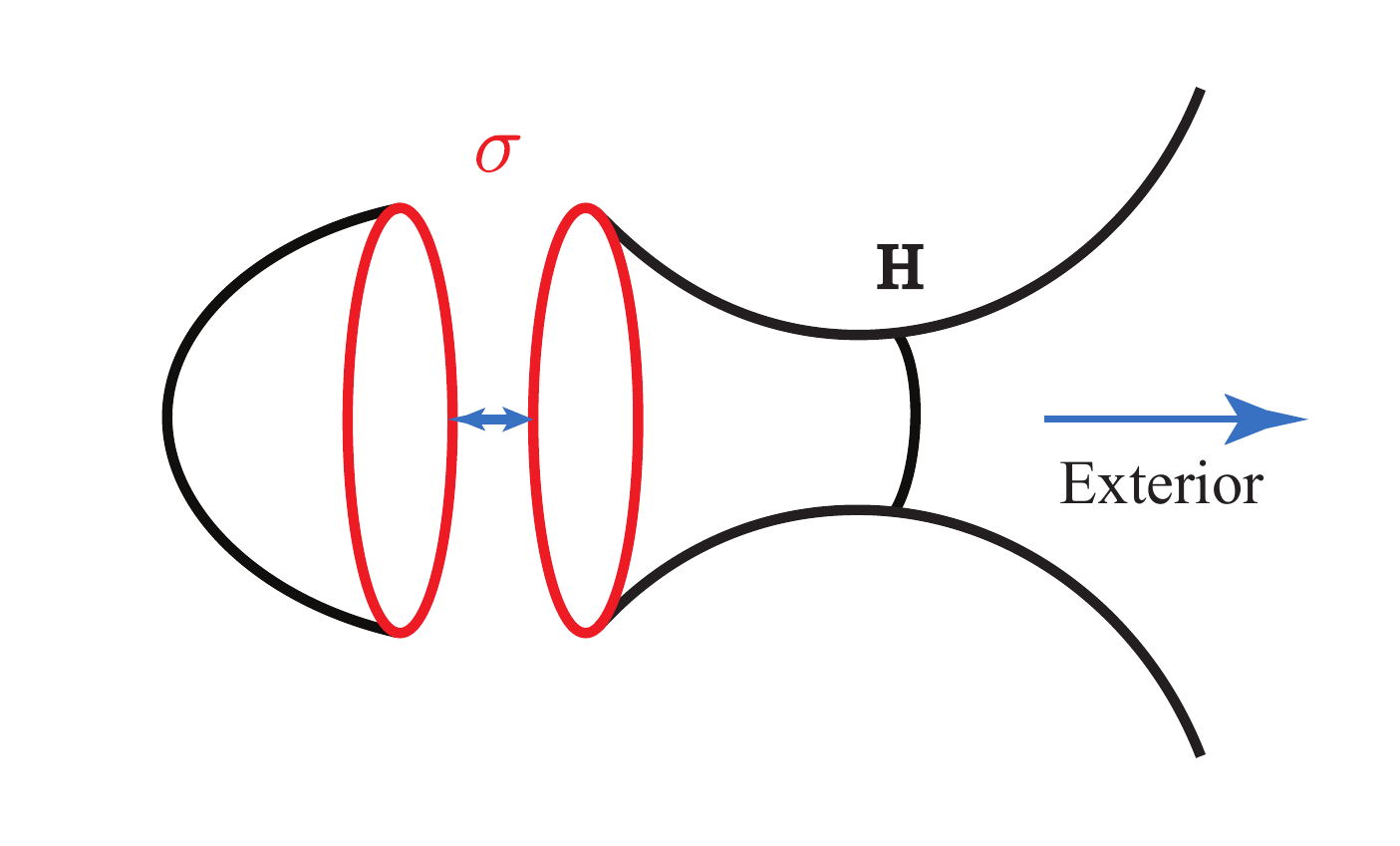}
    \caption{Induced geometry at the time-reflection symmetric slice $\Sigma$. The horizon is at $\mathbf{H} = \mathscr{H}^+ \cap \mathscr{H}^-$. The maximum surface $\sigma = \Sigma \cap \mathcal{W}$ represents the position of the shell, at a radius $R_*\geq r_s$ inside the black hole. The geometry inside the shell caps of smoothly, and it is a portion of flat spacetime.}
    \label{fig:shell2}
\end{figure}

We start by constructing  an infinite family of microstates for an eternal, asymptotically flat (Minkowski), one-sided black hole (Fig.~\ref{fig:shell}). By microstate we refer to a quantum state in the fundamental Hilbert space of the black hole, with fixed values of the coarse-grained observables, such as the mass of the black hole. The microstates that we will consider will have effective semiclassical descriptions, in terms of black hole solutions of general relativity coupled to matter. In these solutions, the black hole is not formed from collapse -- rather, it exists forever, and behind the event horizon there is a ``white hole'' singularity where time begins,  in addition to a black hole singularity where time ends.  However, as we will discuss below, their geometry matches the late-time behavior of black holes forming from collapse, and they can account for the entropy associated to the collapsing configuration. All the states we construct  have the same  geometry between the horizon and the asymptotic spacetime boundary; as such they are microstates of the same black hole as seen by an external observer. Typical black hole microstates are expected to contain Planck scale structures and other features making a geometric description difficult. By contrast, our atypical microstates have a completely well-defined interior: at any time the geometry of  space behind the horizon caps off smoothly (Fig.~\ref{fig:shell2}).  Previously, ``bag of gold'' geometries of this kind, originally described by Wheeler \cite{Wheeler}, were thought to present a conceptual problem because they naively overcount the microstates of black holes.  We will  see how these {\it atypical}\ bags of gold can in fact account precisely for the black hole entropy.

More precisely, outside the horizon all our microstates match the geometry of a Schwarzschild black hole of radius $r_s = 2GM$, where $M$ is the ADM mass. They are distinguished by their interior geometries: each contains a different configuration of matter which backreacts to generate a distinct interior. The matter emerges out of the past singularity and dives into the future singularity, without leaving the black hole region (Fig.~\ref{fig:shell}). For simplicity, we restrict to matter organized in spherical thin shells of dust particles, with total rest mass $m$. The states in the family are labelled by the mass $m$ of the shell in the interior. 

In detail, the exterior metric is the usual Schwarschild one:
\begin{equation}\label{eq:schwarzschild}
\text{d}s^2 = -f(r) \,\text{d}t^2 + \dfrac{\text{d}r^2}{f(r)} + r^2\text{d}\Omega^2\;,
\end{equation}
where $f(r) = 1-\frac{r_s}{r}$ and $\text{d}\Omega^2 = d\theta^2 + \sin^2 \theta \,\text{d}\varphi^ 2$ is the round metric of the unit sphere $\mathbf{S}^2$.

This metric can be continued into the interior of the black hole, at $r < r_s$; and further to a second asymptotic region, described by the same metric (\ref{eq:schwarzschild}). 
Our microstates correspond to shells of matter which live in the black hole interior and this second asymptotic region. In the thin-shell limit, the full geometry inside and outside the shell is determined by the Israel junction conditions \cite{Israel:1966} which fixes the  change in the spacetime metric across the shell.  Concretely, the worldvolume $\mathcal{W}$ of the thin shell carries the localized energy-momentum of a pressureless perfect fluid
\begin{equation}
T_{\mu\nu}|_{\mathcal{W}} = \sigma u_\mu u_\nu \;,
\end{equation}
where $\sigma$ is the surface density of the fluid, and $u^\mu$ is the four-velocity field of the dust, tangent to $\mathcal{W}$. The induced metric on the worldvolume $\mathcal{W}$ is determined by $R(T)$,  the radius of the shell $R$ as a function of its proper time $T$. From the point of view of the metric (\ref{eq:schwarzschild}), the shell will live at $r = R(T)$, with $T$ determined by the proper time along the shell's trajectory.  The equation of motion for $R(T)$, determined by the Israel junction conditions, is that of a non-relativistic particle of zero total energy
\begin{equation}
\dot{R}^2 + V_{\text{eff}}(R) = 0\;,
\end{equation}
where we defined the effective potential
\begin{equation}
V_{\text{eff}}(R) = f(R) - \left(\dfrac{M}{m} - \dfrac{Gm}{2R}\right)^2\;.
\end{equation}
Provided that $M\leq m$, the shell will expand from the past singularity, located at finite proper time in the past, and enter the second exterior region, where it reaches a maximum radius $R_* \geq r_s$ at which $V_{\text{eff}}(R_*) = 0$. The shell then turns around and dives into the future interior and finally the future singularity. For large proper mass, $m \gg M$, the shell re-collapses at a radius $R_* \approx Gm/2$, due entirely to its gravitational self-energy.

The geometry inside the shell consists of a portion of flat space 
\begin{equation}\label{eq:flatinterior}
\text{d}s_<^2 = -\text{d}{\tilde t}^2 + \text{d}{\tilde r}^2 + {\tilde r}^2\text{d}\Omega^2\;,    
\end{equation}
which caps of smoothly at ${\tilde r}=0$. The geometry outside the shell is given by a two-sided black hole geometry, cut off at $r=R_*\geq r_s$ on the left side.

To better understand this geometry, we can focus on the time-reflection symmetric hypersurface $\Sigma$ (Fig.~\ref{fig:shell}) on which we can define the microstate via Euclidean saddle point methods. The induced geometry of this slice resembles a Wheeler ``bag of gold''  \cite{Wheeler}. When the shell mass is large  $m\gg M$, the total interior volume of $\Sigma_{\text{in}} \subset \Sigma$ scales as $\text{Vol}(\Sigma_{\text{in}}) \approx  \frac{\pi}{3}(Gm)^3$, while the surface of the shell $\sigma = \Sigma \cap \mathcal{W}$ has maximum area scaling as $\text{Area}(\sigma) \approx \pi (Gm)^2$ (see Fig.~\ref{fig:shell2}).

The states we have constructed are labeled by the mass of the interior shell. A simple generalization is to consider multiple shells in the interior. We could also consider other configurations of  matter, including exotic matter arising from  string theory.  This choices will generate different interior geometries, while keeping the exterior fixed. As we will argue below, none of these details will matter for counting the microstates. 

The fact that the objects that we construct have horizons and semiclassical interiors, like the conventional black hole, is compatible with the fact that they should in fact be regarded as black hole microstates. In our picture, the horizon is a coarse-grained notion. Thus, a microstate can appear to have a horizon if we probe it with coarse-grained observables. This is entirely in analogy with a gas of particles in a room. The gas can been in a definite microstate, but coarse probes will interact with it as if the system has an entropy. This is because all of the microstates have the same macroscopic description, and interactions with a probe can move the system easily between all these microstates which have the same energy. In our case, we have devised a family of microstates consisting of shells behind the horizon of different inertial masses, which nevertheless have identical geometries (and hence ADM masses) outside the horizon.  These microstates do have horizons in the semiclassical description, but if an observer outside the black hole were powerful enough to act with fine-grained (i.e., UV-sensitive) probes, they would be able to detect the shell inside. In this sense, the horizon is a coarse-grained notion.

\section{Quantum gravitational overlaps}

\begin{figure}[h]
    \centering
    \includegraphics[width = .5\textwidth]{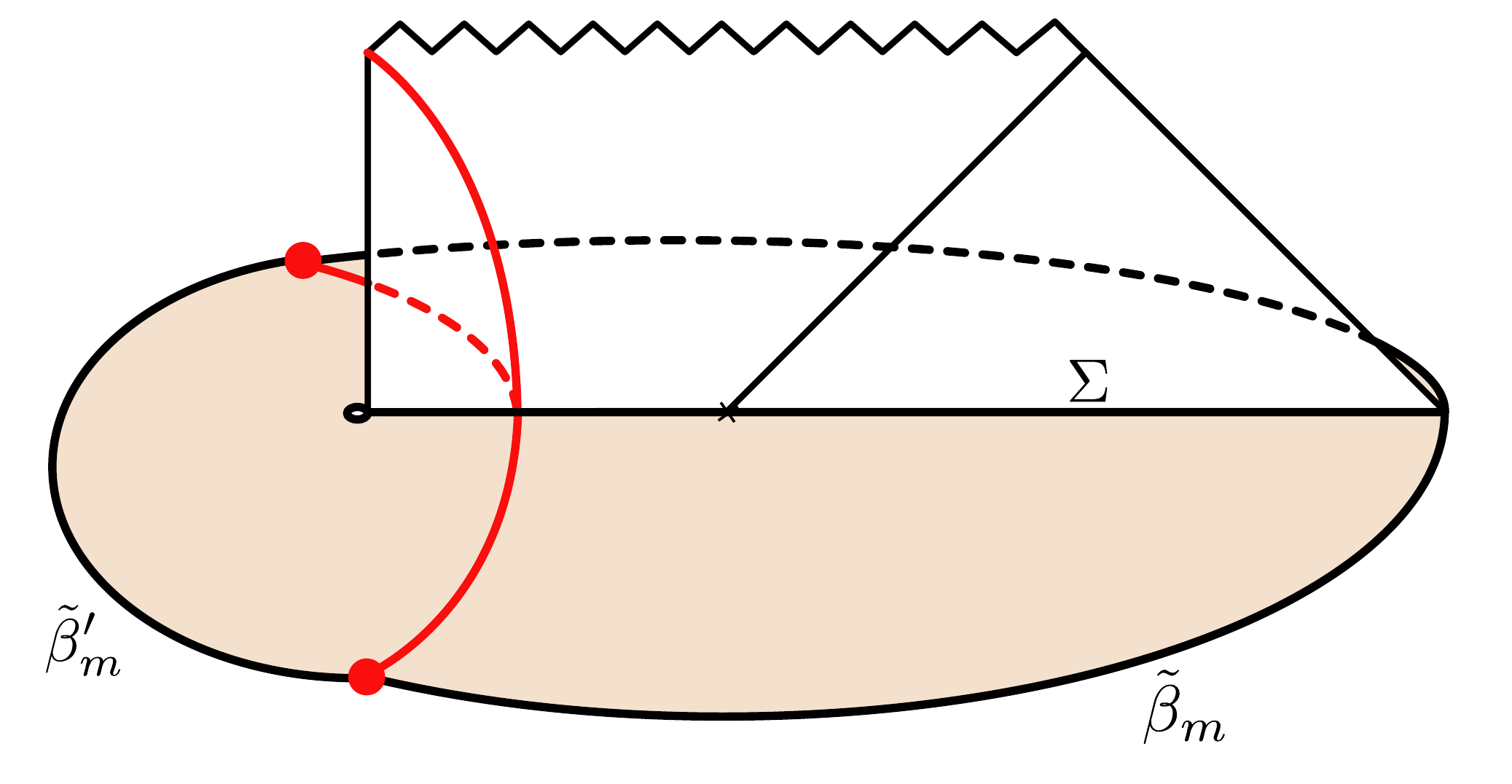}
    \caption{Euclidean continuation of the spacetime geometry of the microstates along $\Sigma$. The Euclidean section consists of an Euclidean black hole (right), and a region of Euclidean flat space (left), glued together along the trajectory of a thin shell. The shell starts at the asymptotic spatial infinity, bounces back at $R_*$ and gets back to $R=\infty$. The euclidean times $\tilde{\beta}_m,\tilde{\beta}_m' \leq \beta$ depend on the mass of the shell.}
    \label{fig:shell3}
\end{figure}

The spacetime geometry $X$ of each microstate can be analytically continued into the Euclidean section along the time reflection-symmetric hypersurface $\Sigma$. The Euclidean geometries define a set of asymptotic boundary conditions at Euclidean spatial infinity $\partial X$. These boundary conditions can be used, within the conventional path integral construction, to prepare the corresponding semiclassical state associated to the microstate of the black hole (see Fig. \ref{fig:shell3}). Related constructions have been studied in models of 2d gravity \cite{Kourkoulou:2017zaj,Goel:2018ubv} and in AdS/CFT \cite{Chandra:2022fwi,Balasubramanian:2022gmo}.

The states constructed in this way are true microstates: they are essentially quantum balls of dust.  Of course such objects, as well as black holes, exist as low energy quantum states in any reasonable theory of gravity coupled to matter, and may have even more microscopic descriptions in terms of strings or D-branes.  Whatever the operators are that produce the particles/strings, there is an explicit path integral construction of our microstates where we apply the operator to the distant past of the Euclidean path integral and evolve forward in Euclidean time. The semiclassical description of these states may be shared  by many string-scale or Planck-scale microstates -- moving a microscopic particle by a Planck length makes little semiclassical difference.

The semiclassical description of the microstates are constructed on the time reflection symmetric ($t=0$) slice $\Sigma$ via the Euclidean path integral in which  there are no singularities. In this way, the construction of the state, and the slice where it lives, are both perfectly regular, analogously to the preparation of the well-known  Hartle-Hawking state in the eternal black hole.  Our state then defines  regular initial data  determining Lorentzian time evolution into the future and the past. The semiclassical description of this time evolution is singular in the black hole interior in the late future and past, but the singularities are an artifact of the breakdown of the semiclassical description and will be absent in the UV-complete description.
Fortunately, we do not need time evolution anywhere in the paper, since we can extract the relevant information from the smooth Euclidean path integral description of the state at $t=0$.

In this way, the microstates that we have discussed in the previous section specify an infinite family of quantum states $\lbrace \ket{\Psi_m} \rbrace$ of the Hilbert space of the black hole, where we remind that $m$ labels the proper mass of the corresponding matter insertion in the black hole interior, and that there is no upper bound on $m$.

This infinite family naively overcounts the Bekenstein-Hawking entropy. But this is only so if the states are orthogonal to each other. To get a correct counting of the dimension of the Hilbert space of the black hole, we must compute the overlaps $\bra{\Psi_m}\ket{\Psi_{m'}}$ between our microstates. This can be done using the gravitational path integral. The rules are to fix the asymptotic boundary conditions that prepare the respective states, and to fill in the Euclidean geometry with all possible saddlepoint manifolds that respect these boundary conditions. We will work in the simple effective description of the microstates, in terms of the Euclidean gravitational action coupled to a thin shell
\begin{gather}
I[X]=-\frac{1}{16\pi G}\int_X R + \dfrac{1}{8\pi G} \int_{\partial X} K + \int_{\mathcal{W}} \sigma + I_{\text{ct}} \;.
\end{gather}
Here $R$ is the Ricci scalar of the Euclidean manifold $X$, $K$ is the extrinsic curvature of its boundary $\partial X$, $\sigma$ is the density of the shell, ${\cal W}$ is the worldvolume of the shell, and  $I_{\text{ct}}$ is a background substraction counterterm that removes divergences and renormalizes the value of the on-shell action.

The leading contribution to the overlap comes from the Euclidean manifold in Fig.~\ref{fig:shell3} which has a single asymptotic boundary where the shell trajectory starts and ends. When $m = m'$ these are straightforward to construct and are simple analogs of those consruced in \cite{Balasubramanian:2022gmo,Sasieta:2022ksu}. When $m \neq m'$, we need some way to join the shells from he Eucliean boundary. This will come from interactions and we expect the result to be exponentially suppressed in the mass difference. We take this difference to be arbitrarily large, so that at leading order the overlap is:
\begin{equation}\label{eq:overlap1}
\overline{\bra{\Psi_m}\ket{\Psi_{m'}}} = \delta_{mm'}\;,
\end{equation}
where the overbar denotes that the calculation is performed according to the rules of the gravitational path integral, and where implicitly we have normalized the states using the on-shell action of the corresponding euclidean manifold $Z_1^{(m)} = e^{-I[X_m]}$. Therefore, microstates representing different classical geometries naively appear to be orthogonal, suggesting our family of microstates actually spans an infinite dimensional Hilbert space.

\begin{figure}[h]
    \centering
    \includegraphics[width = .18\textwidth]{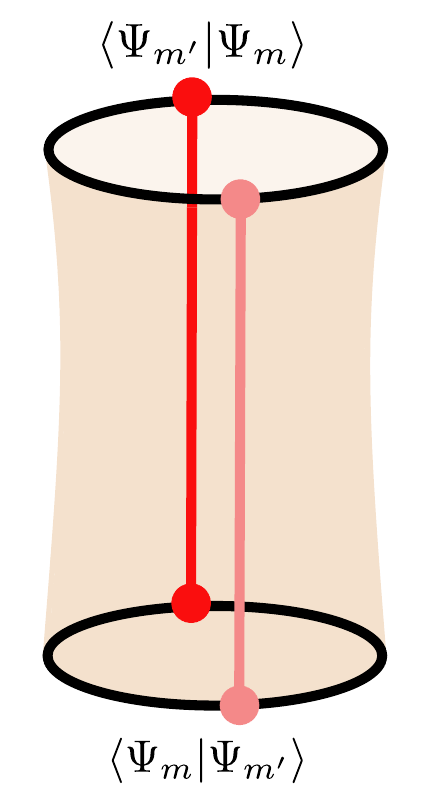}
    \caption{Euclidean wormhole contribution to the second moment of the overlap. The wormhole has the two inner products as its boundaries. It consists in two euclidean black holes in flat space, glued along the two shells.}
    \label{fig:wormhole}
\end{figure}

However, this conclusion is drastically changed by the appearance of semiclassical wormhole contributions in the Euclidean path integral that compute higher moments of this type of amplitudes. These wormholes correspond to non-perturbative effects in quantum gravity. To start with, an explicit two boundary wormhole contributes to the square of the overlap
\begin{equation}\label{amps}
\overline{\left|\bra{\Psi_m}\ket{\Psi_{m'}}\right|^2} = \delta_{mm'} + \dfrac{Z_2}{Z_1^{(m)} Z_1^{(m')}}\;.
\end{equation}
The new contribution is given by the action $Z_2  = e^{-I[X_2]}$, where $X_2$ is the Euclidean wormhole manifold that extends between the two asymptotic boundaries that prepare each of the overlaps (see Fig. \ref{fig:wormhole}).
These give order ${\cal O}(e^{-S})$ contributions which will dominate any mass-dependent terms in the no-wormhole contribution. Specifically, the wormhole solution is constructed by cutting and gluing two Euclidean black hole solutions along the trajectories of the two thin shells. They are the only fully connected solutions with multiple boundaries because the thin shells can only propagate freely in the Euclidean geometry, as they correspond to classical heavy matter. Therefore, each shell has to start at one asymptotic boundary and end on another. Given the fixed boundary points, there is a single trajectory for each shell determined by the equations of motion, and thus there is a single fully connected solution. The detailed construction of such wormholes can be found in \cite{Chandra:2022fwi,Sasieta:2022ksu} for the case of AdS space, and we extend them here to the case of Minkowski spacetime. In the construction of these wormholes, the presence of the shells is crucial to obtain the connected overlap from a semiclassical wormhole.

Similarly, the connected contribution to the $n$-th product is non-vanishing due to the appearence of n-boundary wormholes 
\begin{equation}\label{eq:overlap}
\overline{\bra{\Psi_{m_1}} \ket{\Psi_{m_2}}\bra{\Psi_{m_2}} \ket{\Psi_{m_3}}...\bra{\Psi_{m_n}} \ket{\Psi_{m_1}}}|_c =\dfrac{Z_n}{Z^{(m_1)}_1...\,Z^{(m_n)}_1}\,.
\end{equation}
The connected contribution $Z_n = e^{-I[X_n]}$ corresponds to a semiclassical wormhole $X_n$ with $n$ boundaries. Again, this wormhole is a classical solution to equations of motion. It consists in two black holes joined through the different shells, see \cite{Sasieta:2022ksu,Balasubramanian:2022gmo} for explicit details in AdS.

From the Lorentzian perspective it may seem surprising that data of a microstate hidden behind a horizon enter into action computations that are usually sensitive to asymptotic data. However, in the Euclidean manifold that we use to prepare the semiclassical states (Fig. \ref{fig:shell3}), there is no notion of interior or exterior, since there is no real time-evolution and thus no horizon. A particular microstate is determined by the asymptotic data at the Euclidean boundary of the gravitational path integral. In our case, this data includes the thermal boundary condition, together with the addition of the microscopic operator which creates the shell. Thus, in the calculation of the Euclidean action there is no obstacle for the data of the shell to enter the value of action.

The ``non-factorization'' of the inner products due to non-perturbative effects in quantum gravity might also seem disturbing at first sight. We provide a simple microscopic interpretation of these overlaps and non-factorization in the Appendix. The interpretation is based on the Eigenstate Thermalization Hypothesis \cite{PhysRevA.43.2046,Srednicki_1994}. Briefly, applied here it asserts that these amplitudes should be viewed as the statistics of the fine grained microstates, whose precise computation would involve the control of erratic phases. These erratic phases are invisible to the gravity computation which naturally treat them as random and performs an effective average over them. Non-factorization is naturally associated to an average over random phases.

Below we will need generic $n$-moments of the inner product. These moments can be computed in a straightforward manner for general shell masses, but the expressions are not very illuminating. Luckily we will only need them in the regime of large masses $m_i \gg M$. In this case, the wormhole action simplifies and reduces to
\begin{equation}\label{Uniov}
\overline{\bra{\Psi_{m_1}} \ket{\Psi_{m_2}}\bra{\Psi_{m_2}} \ket{\Psi_{m_3}}...\bra{\Psi_{m_n}} \ket{\Psi_{m_1}}}|_c \approx \dfrac{Z_{\text{bh}}(n\beta)}{(Z_{\text{bh}}(\beta))^n}\;.
\end{equation}
where $Z_\text{bh}(\beta) = e^{-I_\text{bh}(\beta)}$ and $I_\text{bh}(\beta)$ is the gravitational action of the Euclidean Schwarzschild black hole of inverse temperature $\beta$, where $\beta = 8\pi G M$ is the inverse temperature of the original black hole. Equivalently, the action $I_\text{bh}(\beta)$ is the euclidean gravity action first computed in the seminal article by Gibbons and Hawking \cite{Gibbons:1977}.

Notice that these results provide a new interpretation of the meaning of the Gibbons-Hawking action.  This quantity  appears here in the  context of  the construction of microstates and their overlaps. By contrast, Gibbons and Hawking were computing the free energy of a thermal state. In particular, the asymptotic boundary conditions that we impose for the Euclidean path integral includes thin shell operator insertions, which is why our gravitational solutions compute overlaps.

Hence we conclude that the overlaps become universal in this limit, independent of the actual masses of the shells characterizing the microstates. The physical intuition is that these microstates become relatively random in this limit. The average overlap between two random states is universal and only depends on the Hilbert space dimension that these states live in. As we will see in the next section, these universal overlaps encode the dimension of the real Hilbert space of the black hole.

\section{Counting microstates}

We now consider an infinite subfamily of black hole microstates $\lbrace \ket{\Psi_{m_j}}\rbrace$ for shells with mass $m_j = j\,m$ for $j = 1,2,...$, where $m$ is a sufficiently large value of the mass. The objective is to determine the dimension of the Hilbert space spanned by these microstates. To this end we consider the Gram matrix of overlaps for a finite subset of these states
\begin{equation}
G_{ij} = \bra{\Psi_{m_i}} \ket{\Psi_{m_j}} \;,
\end{equation}
where $i,j = 1,...,\Omega$. This Gram matrix is hermitian and positive semidefinite. The number of zero eigenvalues of this matrix counts the number of linearly independent microstates in the given subset. Equivalently, the rank of this matrix gives the Hilbert space dimension spanned by the set. We seek to compute the rank of the Gram matrix as a function of $\Omega$.

As mentioned above, the gravitational computations of the moments of this Gram matrix suggest that we interpret it as a random matrix with moments given by the universal overlaps~(\ref{Uniov}). In order to count black hole microstates for a given energy from these overlaps, i.e. the microcanonical degeneracy, we project the previous universal expressions into a given energy via an inverse Laplace transform of the wormhole contributions
\begin{equation}
Z_{\text{bh}}(n\beta)= \int dE \,\rho(E)\,e^{-n\beta E} \;,
\end{equation}
and for a given microcanonical window of energies $[E,E+\Delta E]$, we define the functions
\begin{equation}\label{micro}
   \;\;\;\;\;\;\; e^\textbf{S}\equiv\,\rho(E)\,\Delta E\,\;,\,\,\,\,\,\,\,\,\,\,\,\,\,\,\,\,\,\,\,\,\,\textbf{Z}_n\equiv\,\rho(E)\,e^{-n\beta\,E}\,\Delta E\;.
\end{equation}
The function $\mathbf{S}$ coincides with the Bekenstein-Hawking entropy \cite{Bekenstein:1973ur,Hawking:1975vcx}
\begin{equation}\label{bhent}
\mathbf{S} = \dfrac{A}{4G}\;,
\end{equation}
but notice that, at this stage, this function lacks the interpretation in terms of the dimension of black hole Hilbert space. Here it is just re-contextualized as a function controlling non-perturbative wormhole contributions to the gravitation path integral.

Using the microcanonical expressions (\ref{micro}), in the Appendix we show that the density of states of the Gram matrix is given by
\begin{gather}
D(\lambda)=\frac{e^\textbf{S}}{2\pi\lambda}\sqrt{\left[\lambda-\left(1-\sqrt{\frac{\Omega}{ e^{\textbf{S}}}}\right)^2\,\right]\left[\,\left(1+\sqrt{\frac{\Omega}{ e^{\textbf{S}}}} \right)^2-\lambda \right]}\nonumber\\+\delta(\lambda)\left(\Omega-e^{\mathbf{S}}\right)\theta(\Omega-e^{\mathbf{S}})\;.
\end{gather}
This density of states has a continuous part and a singular part. The eigenvalues accounted by the continuous part are all positive definite. The singular part counts the number of zero eigenvalues. Therefore, the rank of the Gram matrix is the number of eigenvalues contained in the continuous part of the distribution. We thus conclude:
\begin{itemize}
    \item For $\Omega<e^{\textbf{S}}$, the rank of $G$ is given by $\Omega$.  
    \item For $\Omega>e^{\textbf{S}}$, the rank of $G$ is given by $e^{\textbf{S}}$.
\end{itemize}
For $\Omega < e^{\textbf{S}}$, we can thus use the Gram-Schmidt proceure to construct an orthonormal set of vectors. For $\Omega>e^{\textbf{S}}$ this will fail, as the microstates will no longer be linearly independent. This gives the main result of this article, namely that the black hole microstate degeneracy, equal to the number of possible orthogonal states in a given energy band is equal to the exponential of the Bekenstein-Hawking entropy~(\ref{bhent}). Equivalently, the present microstate construction provides a microscopic statistical understanding of the entropy of black holes in Minkowski spacetime.

These statements are valid on average in the effective ensemble provided by semiclassical gravity. But deviations from the mean in Random Matrix Theory \cite{10.1093/oxfordhb/9780198744191.001.0001}  are suppressed by the dimension of the associated matrix, in this case by factors of $e^{-\textbf{S}}$. It would be interesting to compute these subleading corrections to the Bekenstein-Hawking entropy. However, we note that they will also likely be subleading to other corrections, such as those following from higher-derivative corrections to the gravitational effective action that generically appear in a UV-complete theory.

Let us summarize the intuition behind our result. Even if we keep adding potential microstates, there is a point at which these states cannot be orthogonal anymore. This point is controlled by the universal statistics of the inner product, themselves controlled by the Bekenstein-Hawking entropy. As proposed in the introduction, the solution to the problem of the microscopic origin of the entropy of general black holes is not to construct \emph{a specific set} of $e^{\textbf{S}}$  microstates. Indeed, there are infinite numbers of such sets, even when they are constrained to be semiclassical and geometrical. The problem is to count how many orthogonal states we can build out of those, and prove this counting gives rise to the right Bekenstein-Hawking dimension.

Finally, one could ask if {\it any} black hole microstate is expandable in our basis. Although there might be rare states contributing with subleading corrections to the dimension of the Hilbert space we computed, we indeed expect most states to be expandable in the basis generated from our shell states. There is an explicit procedure to check such question using our techniques. We just compute the inner products between such state and our basis, taking again into account the appearance of wormholes. For example, we could take the Thermofield Double (TFD) state without any shell, along with the small excitations around this background. This space of microstates again has non-zero, exponentially suppressed, overlaps with all shell states that we have considered.  The dominant contribution to these overlaps comes from a wormhole saddle with one shell behind the horizon.  Thus adding the TFD family of states to the Gram matrix will not  increase its rank.

\section{Black holes from collapse}

Gravitational collapse to form black holes can be modeled by a spherical shell of dust falling in from infinity.  The shell, with flat space inside and 
a Schwarzschild geometry outside,  eventually passes behind the spacetime horizon that its own presence generates.   After this passage, classical observers outside the horizon observe a black hole geometry of a fixed mass $M$ in an asymptotically flat universe, but do not know the precise state behind the horizon.  In particular, all the microstates described above for the eternal  black hole lead to same external geometry in the future of the shell.  Thus, by the rules of the microcanonical ensemble, the exterior classical observer will interact with the black hole as if it has an entropy equal to that of an eternal black hole of the same mass.

Most of the quantum states counted by this entropy have pasts that do not resemble the infalling shell.  For example, the interior shells that we used to count the microstates in the previous sections time-evolved from a white hole singularity of the eternal one-sided black hole. For that matter, adding a single interior quantum to a collapsing shell geometry at some time will lead to a state that evolves backwards to having an initial singularity. This is completely consistent with our proposal and with the fundamentals of statisical mechanics. Properly speaking, the entropy of a thermalized system is counted by all of the microstates consistent with the macroscopic parameters after the system has reached thermal equilibrium. Most of the states which contribute to the entropy will have pasts that are very different from the initial conditions in any single process. The pasts of the collapsing shell states are atypical, just as configurations of a gas with all the molecules in one corner of a container are atypical. 

Moreover, we can define quantum states on a  time-slice, for example $\Sigma$ in Fig.~\ref{fig:shell}, without reference to their past under unitary time evolution. At that time, we can still write the collapsing shell geometry in such a basis. The reason is a generalization of arguments given above. This shell state has exponentially small but still non-vanishing overlaps with all states in the basis. These overlaps can be computed using the gravitational path integral. They arise by a wormhole contribution similar to the ones we have been discussing but more difficult to analyze given the time asymmetry of the collapsing shell. This does lead to the unusual statement that quantum superpositions of macroscopic, semiclassical states can yield other macroscopic, semiclassical states. But there is no inconsistency; individual overlaps are extremely small, and we are making use of a very high-dimensional Hilbert space together with a delicate balance of phases.

This suggests a tantalizing scenario. Since collapsing shells are random superposition of interior shells, and since the  collapsing shells leads to a different pasts than the basis of interior shells, these pasts must also be superpositions of each other.  Also, since the interior shell states have a past singularity, while collapsing ones do not, it would appear in a cosmological sense, some universes with a smooth past can also be understood as superpositions of universes with a singular past.  Perhaps it is then possible to go the other way, and show that universes with singular pasts can be understood as quantum superpositions of smooth universes.

\section{Discussion}

The key insight in our approach is that to understand the microscopic origin of black hole entropy it is enough to construct any well-controlled and sufficiently large space of states. The rank of the matrix of overlaps between these states then reveals the Hilbert space dimension.  We have shown that there are many families of such microstates under effective semiclassical control.  These states are highly atypical in the Hilbert space, and do not have the standard Schwarzschild interior. But nevertheless, they are sufficient to demonstrate that the Bekenstein-Hawking entropy can be explained as the dimension of an underlying quantum Hilbert space of the black hole.

Our results highlight that semiclassical gravity has access to some key  non-perturbative features of the microscopic theory, but not to all of them.  The entropy is a coarse-grained but non-perturbative notion that we can access, while other finer properties are washed out in the semiclassical description.  Of course, the black hole entropy will have subleading, non-universal corrections that can be computed explicitly from the semiclassical field content, by including one-loop determinants  correcting the universal overlaps. Still other corrections, related for example to the discretenes of the exact quantum desnity of states, will depend on the precise UV completion.

However, our results do not explicitly use any of these details of string theory, AdS/CFT, or any other UV formulation of quantum gravity.  Indeed, the only assumptions we have used are that: (a) there is some ultraviolet completion, and (b) the semiclassical Euclidean path integral provides sensible information about the ultraviolet completion.  With these assumptions, our results provide an explanation for the entropy of black holes in any theory that has general relativity coupled to massive matter as a low-energy limit. In particular, this construction works in specific top-down models of quantum gravity, such as those appearing in string theory and AdS/CFT, since these theories contain the necessary ingredients to construct our microstates. Perhaps this explains the universality of the Bekenstein-Hawking entropy formula. 
The universality of the overlaps, and thence of the black hole entropy, is an {\it output} of the gravitational path integral that we discover. We propose a  physical interpretation the microstates become in a sense random relative to each other, provided that the masses of their respective shells take very different values. By ``random'' we mean here that act like random superpositions of a fixed basis.  The average overlap between two random states is universal and only depends on the Hilbert space dimension that these states live in.  In a companion paper \cite{Balasubramanian:2022gmo} we repeated the present analysis in universes with a negative cosmological constant because there are additional tools in this case that allow us to put our state-construction methods on an entirely firm footing.

 \section{Acknowledgments}

We would like to thank Jos\'{e} Barb\'{o}n, Alex Belin, Jan de Boer, Horacio Casini, Kristan Jensen, Henry Maxfield and Roberto Emparan for useful discussions.  VB and JM are supported in part by the Department of Energy through  DE-SC0013528 and  QuantISED DE-SC0020360, as well as by the Simons Foundation through the It From Qubit Collaboration (Grant No. 38559). 
AL and MS are supported in part by the Department of Energy through DE-SC0009986 and QuantISED DE-SC0020360. This preprint is assignated the code BRX-TH-6713.

\appendix

\section{Microscopic interpretation of the overlaps}\label{App1}

Inner products between quantum states in a Hilbert space are by definition complex numbers. As such, the product of a collection of them must factorize. However, this is not what the gravitational path integral provides. The presence of wormhole contributions in \eqref{eq:overlap} manifestly spoils factorization. 

This naturally raises a question: what are these overlaps really computing in the fine-grained description of the black hole? In this appendix, we argue that the gravitational path integral computation provides information about the typical magnitude of the overlaps for a collection of quantum states which only differ in fine-grained structural details. The gravitational semiclassical description is insensitive to these fine grained details, but sensitive to the sizes.

The key insight, expanding on \cite{Sasieta:2022ksu}, is to rely on the fact that the Hamiltonian of the black hole is expected to be chaotic (see \cite{Maldacena:2015waa,Cotler:2016fpe} for example), and that the thin shell that lives inside has a simple description in the local effective field theory basis of the black hole interior, i.e. it just creates a collection of particles. The behavior of the matrix elements of such simple operators in chaotic theories is encapsulated in the Eigenstate Thermalization Hypothesis (ETH) \cite{PhysRevA.43.2046,Srednicki_1994}
\begin{equation}   \label{ETHold} 
\bra{E_n} \mathcal{O} \ket{E_m}\,=\,e^{-S(\bar{E})/2)}\, g(\bar{E},\omega)^{1/2} \,R_{nm}\;.
\end{equation}
where the $|E_n\rangle$ are energy eigenstates of the black hole. In this expression we chose an operator with no diagonal part. Equivalently, we chose any operator and substracted the diagonal to create another operator. We also defined $\bar{E} \equiv (E_n+E_m)/2$ and $\omega = E_m - E_n$. The function $g(\bar{E},\omega)$ is smooth in the thermodynamic limit, and it encodes the information about the microcanonical two-point function of the operator. The coefficients $R_{nm}$ are erratic complex numbers of $\mathcal{O}(1)$ magnitude. The ETH then asserts that the $R_{nm}$ entries can be viewed as  independent random variables with zero mean and unit variance. This highly accepted hypothesis naturally provides an effective ensemble of operators. In our case, this effective ensemble induces an ensemble of microstates of the black hole.

We now come back to the moments of the inner products, $\overline{\bra{\Psi_{m_1}} \ket{\Psi_{m_2}}\bra{\Psi_{m_2}} \ket{\Psi_{m_3}}...\bra{\Psi_{m_n}} \ket{\Psi_{m_1}}}$.
In this expression, through the article the overline has been used to denote a computation based on the gravity path integral. By writing every state in terms of the shell operators used to prepare them, and assuming ETH for the thin shell, we can interpret the overline in terms of an ETH average of the form
\begin{equation}  \label{CorrETH} 
\overline{\mathcal{O}_{n_1m_2}\,\mathcal{O}_{n_2m_3}\,\cdots\,\mathcal{O}_{n_km_1}}\;,
\end{equation}
where $\mathcal{O}_{nm} = \bra{E_n} \mathcal{O} \ket{E_m}$.  In this way we arrive at a very simple interpretation of the non-factorization of the gravitational inner products for our microstates: the semiclassical path integral only computes averages over the ETH ensemble of operators. The fine-grained phase of the overlap $\bra{\Psi_{m}} \ket{\Psi_{m'}}$ for different masses depends erratically on the ETH coefficients of the operators $\mathcal{O}_m$ and $\mathcal{O}_{m'}$. It then averages out to zero semiclassically,  $\overline{\bra{\Psi_{m}} \ket{\Psi_{m'}}}=0$. On the contrary, the magnitude of the overlap is a self-averaging quantity, and its typical value is captured by the wormhole contribution in \eqref{amps}. Equivalently it is captured by the exponentially small but still non-vanishing ETH correlations~(\ref{CorrETH}).

\section{Density of states from Schwinger-Dyson equation}\label{App2}

As mentioned above the Gram matrix of microstates should be interpreted as a random matrix. The n-moment of this random matrix is computed by the gravitational path integral and reads
\begin{equation}\label{overap}
\overline{G_{i_1 i_2}\,G_{i_2 i_3}\,\cdots\,G_{i_n i_1}}|_c \approx \dfrac{Z_{\text{bh}}(n\beta)}{(Z_{\text{bh}}(\beta))^n}\;.
\end{equation}
In this appendix we review how to compute the density of states in this scenario. We start by writing down the resolvent $R$ of the Gram matrix $G$, defined by
\begin{equation}
R_{ij}(\lambda)\,\equiv\,\left( \frac{1}{\lambda \mathds{1} -G}\right)_{ij} \,=\,\frac{1}{\lambda}\,\delta_{ij}+\sum\limits_{n=1}^{\infty}\,\frac{1}{\lambda^{n+1}}\,(G^n)_{ij}  \;.
\end{equation}
Using the gravitational path integral, we are able to compute the resolvent $\overline{R_{ij}(\lambda)}$ in terms of the gravitational overlaps $\overline{(G^n)_{ij}}$ given by the previous equation~(\ref{overap}). For these types of matrix taylor expansions, as shown in \cite{Penington:2019kki}, see also \cite{Balasubramanian:2022gmo}, one can write down a Schwinger-Dyson equation for the resolvent 
\begin{equation}\label{eq:sdresolvent}
\overline{R_{ij}(\lambda)}=\frac{1}{\lambda}\,\delta_{ij}+\frac{1}{\lambda}\,\sum\limits_{n=1}^{\infty}\,\frac{Z(n\beta)}{Z(\beta)^n}\,\overline{R(\lambda)}^{\,n-1}\,\overline{R_{ij}(\lambda)}\;,
\end{equation}
where $R(\lambda) = \sum_{i=1}^\Omega R_{ii}(\lambda)$ is the trace. 

We now project the previous moments into a given energy window by doing the inverse Laplace transform
\begin{equation}
Z_{\text{bh}}(n\beta) = \int dE \,\rho(E)\,e^{-n\beta E} \;.
\end{equation}
Focusing on the corresponding microcanonical window of energies $[E,E+\Delta E]$, we define
\begin{equation}
   \;\;\;\;\;\;\; e^\textbf{S}\equiv\,\rho(E)\,\Delta E\,\;,\,\,\,\,\,\,\,\,\,\,\,\,\,\,\,\,\,\,\,\,\,\textbf{Z}_n\equiv\,\rho(E)\,e^{-n\beta\,E}\,\Delta E\;.
\end{equation}
Taking the trace on both sides of \eqref{eq:sdresolvent} and inserting $\textbf{Z}_n$ instead of $Z(n\beta)$ in the Schwinger-Dyson equation makes it possible to perform the sum and write down a quadratic equation for the trace of the resolvent
\begin{equation}\label{solr}
\overline{R(\lambda)}^2+\left(\,\frac{e^\textbf{S}-\Omega}{\lambda}-e^\textbf{S}\,\right)\,\overline{R(\lambda)}+\Omega\,e^\textbf{S}=0\;.
\end{equation}

Given the definition of the resolvent, the density of states of the Gram matrix $G$ follows from the discontinuity across the real axis of the resolvent $R$. More precisely
\begin{equation}
D(\lambda)=\frac{1}{2\,\pi\,i}\left(\,\overline{R(\lambda-i\epsilon)}-\overline{R(\lambda+i\epsilon)}\,\right)\;.
\end{equation}
Using the algebraic solution to~(\ref{solr}) we finally arrive at the density of states used previously in the article
\begin{gather}
D(\lambda)=\frac{e^\textbf{S}}{2\pi\lambda}\sqrt{\left[\lambda-\left(1-\sqrt{\frac{\Omega}{ e^{\textbf{S}}}}\right)^2\,\right]\left[\,\left(1+\sqrt{\frac{\Omega}{ e^{\textbf{S}}}} \right)^2-\lambda \right]}\nonumber\\+\delta(\lambda)\left(\Omega-e^{\mathbf{S}}\right)\theta(\Omega-e^{\mathbf{S}})\;.
\end{gather}


\begin{thebibliography}{17}

        \bibitem{Bekenstein:1973ur}
        J.~D.~Bekenstein,
        ``Black holes and entropy",
        Phys. Rev. D \textbf{7}, 2333-2346 (1973)
        doi:10.1103/PhysRevD.7.2333

        \bibitem{Hawking:1975vcx}
        S.~W.~Hawking,
        ``Particle Creation by Black Holes",
        Commun. Math. Phys. \textbf{43}, 199-220 (1975)
        [erratum: Commun. Math. Phys. \textbf{46}, 206 (1976)]
        doi:10.1007/BF02345020

        \bibitem{Strominger:1996sh}
        A.~Strominger and C.~Vafa,
        ``Microscopic origin of the Bekenstein-Hawking entropy",
        Phys. Lett. B \textbf{379}, 99-104 (1996)
        doi:10.1016/0370-2693(96)00345-0
        [arXiv:hep-th/9601029 [hep-th]].


        \bibitem{Balasubramanian:2022gmo}
        V.~Balasubramanian, A.~Lawrence, J.~M.~Magan and M.~Sasieta,
        ``Microscopic origin of the entropy of black holes in general relativity'',
        [arXiv:2212.02447 [hep-th]].

        

        \bibitem{Israel:1966} 
        W.~Israel, ``Singular hypersurfaces and thin shells in general relativity", {Nuovo Cimento B Serie} {\bf 44} 1. (1966) doi:10.1007/BF02710419.

        \bibitem{Wheeler} 
        J.~Wheeler,
        ``Relativity, groups and topology", edited by B. S. DeWitt and C. M. DeWitt. Gordon and Breach New York (1964).



     
       \bibitem{Kourkoulou:2017zaj}
        Kourkoulou, Ioanna and Maldacena, Juan,
        ``Pure states in the SYK model and nearly-$AdS_2$ gravity'',
        [arXiv:1707.02325 [hep-th]].



       
       
       
       
       
       \bibitem{Goel:2018ubv}
        Goel, Akash and Lam, Ho Tat and Turiaci, Gustavo J. and Verlinde, Herman,
        ``Expanding the Black Hole Interior: Partially Entangled Thermal States in SYK'',
        JHEP \textbf{02}, 156 (2019)
        [arXiv:1807.03916 [hep-th]].









        

   
        \bibitem{Chandra:2022fwi}
        J.~Chandra and T.~Hartman,
        ``Coarse graining pure states in AdS/CFT'',
        [arXiv:2206.03414 [hep-th]].
       


        
        \bibitem{Sasieta:2022ksu}
        M.~Sasieta,
        ``Wormholes from heavy operator statistics in AdS/CFT,''
        JHEP \textbf{03}, 158 (2023)
        doi:10.1007/JHEP03(2023)158
        [arXiv:2211.11794 [hep-th]].




      \bibitem{PhysRevA.43.2046}
        Deutsch, J. M.,
        ``Quantum statistical mechanics in a closed system",
        Phys. Rev. A, \textbf{43}, 4 (1991).




      \bibitem{Srednicki_1994}
        Mark Srednicki,
        ``Chaos and quantum thermalization",
        Phys. Rev. E, \textbf{50}, 2 (1994).


        \bibitem{Penington:2019kki}
        G.~Penington, S.~H.~Shenker, D.~Stanford and Z.~Yang,
        ``Replica wormholes and the black hole interior'',
        JHEP \textbf{03}, 205 (2022)
        doi:10.1007/JHEP03(2022)205
        [arXiv:1911.11977 [hep-th]].

        \bibitem{Gibbons:1977}
        G. W. ~Gibbons and S. W. ~Hawking,
        ``Action integrals and partition functions in quantum gravity",
        Phys. Rev. D, \textbf{15}, 2752--2756 (1977)
         doi:10.1103/PhysRevD.15.2752
        [https://link.aps.org/doi/10.1103/PhysRevD.15.2752]

        
        

\bibitem{10.1093/oxfordhb/9780198744191.001.0001}
        Akemann, Gernot and Baik, Jinho and Di Francesco, Philippe,
        ``The Oxford Handbook of Random Matrix Theory",
        Oxford University Press \textbf{09} 2015




\bibitem{Maldacena:2015waa}
J.~Maldacena, S.~H.~Shenker and D.~Stanford,
``A bound on chaos,''
JHEP \textbf{08} (2016), 106
[arXiv:1503.01409 [hep-th]].



    \bibitem{Cotler:2016fpe}
        Cotler, Jordan S. and Gur-Ari, Guy and Hanada, Masanori and Polchinski, Joseph and Saad, Phil and Shenker, Stephen H. and Stanford, Douglas and Streicher, Alexandre and Tezuka, Masaki,
        ``Black Holes and Random Matrices'',
        JHEP \textbf{05}, 118 (2017)
        doi:10.1007/JHEP05(2017)118
        [arXiv:1611.04650 [hep-th]].

 








    
\end{thebibliography}
\end{document}